\documentclass[conference,letterpaper]{IEEEtran}
\usepackage[letterpaper, left=0.69in, right=0.69in, bottom=1.05in, top=0.65in]{geometry}
\IEEEoverridecommandlockouts

\usepackage{cite}
\usepackage{amsmath,amssymb,amsfonts, mathtools, amsthm}
\usepackage{algorithmic}
\usepackage{algorithm}
\usepackage{lipsum}
\usepackage{subcaption}
\usepackage{epsfig}
\usepackage{bbm}
\usepackage{bm}
\usepackage{booktabs}
\usepackage{multirow}
\usepackage[normalem]{ulem}
\usepackage{graphicx, subfig}
\usepackage{textcomp}
\usepackage{xcolor}
\usepackage{verbatim}
\usepackage{glossaries}
\usepackage{threeparttable}
\usepackage{tabularx}
\usepackage{caption}
\usepackage{dsfont}
\usepackage{titlesec}
\usepackage[bb=dsserif]{mathalpha}
\usepackage{pgfplots}
\usepackage{comment}

\newacronym{bcd}{BCD}{Block Coordinate Descent}
\newacronym{leo}{LEO}{Low Earth Orbit}
\newacronym{isl}{ISL}{Inter-Satellite Link}
\newacronym{gsd}{GSD}{ground sample distance}
\newacronym{fov}{FoV}{field of view}
\newacronym{gtfp}{GTFP}{ground track frame period}
\newacronym{gs}{GS}{ground station}
\newacronym{fso}{FSO}{free-space optical}
\newacronym{smec}{SMEC}{satellite mobile edge computing}
\newacronym{mec}{MEC}{mobile edge computing}
\newacronym{pdd}{PDD}{Penalty Dual Decomposition}
\newacronym{aodv}{AODV}{Ad hoc On-demand Distance Vector}
\newacronym{cnn}{CNN}{convolutional neural networks}
\newacronym{dl}{DL}{Deep Learning}
\newacronym{dod}{DoD}{depth of discharge}
\newacronym{dqn}{DQN}{deep Q-learning}
\newacronym{gsl}{GSL}{ground-to-satellite link}
\newacronym{bm}{BM}{benchmark}
\newacronym{ml}{ML}{Machine Learning}
\newacronym{mdp}{MDP}{Markov decision process}
\newacronym{ngeo}{NGEO}{Non-geostationary orbit}
\newacronym{olsr}{OLSR}{optimized link state routing protocol}
\newacronym{ospf}{OSPF}{Open Shortest Path First}
\newacronym{pan}{PAN}{Path-Aware Networking}
\newacronym{qos}{QoS}{Quality of Service}
\newacronym{rl}{RL}{Reinforcement Learning}
\newacronym{drl}{DRL}{Deep Reinforcement Learning}
\newacronym{dnn}{DNN}{Deep Neural Network}
\newacronym{dql}{DQL}{Deep Q-learning}
\newacronym{e2e}{E2E}{end-to-end}
\newacronym{bgp}{BGP}{Border Gateway Protocol}
\newacronym{ibgp}{iBGP}{interior Border Gateway Protocol}
\newacronym{ebgp}{eBGP}{exterior Border Gateway Protocol}
\newacronym{as}{AS}{Autonomous System}
\newacronym{relu}{ReLu}{Rectified Linear Unit}
\newacronym{cdf}{CDF}{Cumulative Distribution Function}
\newacronym{ntn}{NTN}{Non-Terrestrial Networks}
\newacronym{lsatc}{LSatC}{\gls{leo} Satellite Constellation}
\newacronym{ai}{AI}{Artifical Intelligence}
\newacronym{ip}{IP}{Internet Protocol}
\newacronym{ue}{UE}{User Equipment}
\newacronym{pomdp}{POMDP}{Partially Observable Markov Decision Problem}
\newacronym{hol}{HOL}{Head of Line}
\newacronym{fifo}{FIFO}{First-In First-Out}
\newacronym{snr}{SNR}{Signal-to-Noise Ratio}
\newacronym{eo}{EO}{Earth Observation}
\newacronym{aoi}{AoI}{Age of Information}
\newacronym{paoi}{PAoI}{Peak Age of Information}
\newacronym{semcom}{SemCom}{Semantic Communications}
\newacronym{go}{GO}{Goal-Oriented Communications}
\newacronym{vtw}{VTW}{Visible Time Window}
\newacronym{otw}{OTW}{Observation Time Window}
\newacronym{aeossp}{AEOSSP}{Agile Earth Observation Satellite Scheduling Problem}
\newacronym{aeos}{AEOS}{Agile Earth Observation Satellite}
\newacronym{ceos}{CEOS}{Conventional Earth Observation Satellite}
\newacronym{ec}{EC}{Edge Computing}
\newacronym{gcn}{GCN}{Graph Convolutional Networks}
\newacronym{gnn}{GNN}{Graph Neural Networks}
\newacronym{gat}{GAT}{Graph Attention Networks}
\newacronym{sgd}{SGD}{stochastic gradient descent}
\newacronym{sth}{STH}{Scheduling Time Horizon}
\newacronym{stp}{STP}{Scheduling Time Period}
\newacronym{oc}{OC}{Opportunity Cost}
\newacronym{ls}{LS}{Local Search}
\newacronym{cpu}{CPU}{Central Processing Unit}
\newacronym{gpu}{GPU}{Graphics Processing Unit}

\def\BibTeX{{\rm B\kern-.05em{\sc i\kern-.025em b}\kern-.08em
    T\kern-.1667em\lower.7ex\hbox{E}\kern-.125emX}}

\title{Scheduling Agile Earth Observation Satellites with Onboard Processing and Real-Time Monitoring} 

\author{\IEEEauthorblockN{Antonio M. Mercado-Martínez, Beatriz Soret~\IEEEmembership{Senior Member,~IEEE}, Antonio Jurado-Navas~\IEEEmembership{Member,~IEEE}}
\vspace{-0.4cm}

\thanks{This work is funded by the Spanish Ministerio de Ciencia, Innovación y Universidades (project "TATOOINE", grant no. PID2022-136269OB-I00). The author thankfully acknowledges the computer resources, technical expertise and assistance provided by the SCBI (Supercomputing and Bioinformatics) center of the University of Malaga.} }

\def\subparagraph{} 

\titlespacing*{\section}{0pt}{*1}{*1}
\titlespacing{\subsection}{0pt}{*1}{*1}

\renewcommand{\thesubsubsection}{\arabic{subsubsection}}

\titleformat{\subsubsection}[runin]{\itshape}{\thesubsubsection)}{1em}{}
\titlespacing*{\subsubsection}{\parindent}{0pt}{*1}
\begin{document}
\bstctlcite{IEEEexample:BSTcontrol}

\maketitle
\begin{abstract}
The emergence of \glspl{aeos} has marked a significant turning point in the field of \gls{eo}, offering enhanced flexibility in data acquisition. Concurrently, advancements in onboard satellite computing and communication technologies have greatly enhanced data compression efficiency, reducing network latency and congestion while supporting near real-time information delivery. In this paper, we address the \gls{aeossp}, which involves determining the optimal sequence of target observations to maximize overall observation profit. Our approach integrates onboard data processing for real-time remote monitoring into the multi-satellite optimization problem. To this end, we define a set of priority indicators and develop a constructive heuristic method, further enhanced with a \gls{ls} strategy. The results show that the proposed algorithm provides high-quality information by increasing the resolution of the collected frames by up to 10\% on average, while reducing the variance in the monitoring frequency of the targets within the instance by up to 83\%, ensuring more up-to-date information across the entire set compared to a \gls{fifo} method.
\end{abstract}

\glsresetall

\section{Introduction}

\gls{eo} is a rapidly evolving and highly relevant application in satellite communications, providing critical insights for climate monitoring, disaster management, maritime surveillance, or vehicle tracking. Traditionally, \gls{eo} has involved the acquisition and transmission of vast volumes of data, posing challenges for both communication and processing. However, the enhanced onboard  computing and communication capabilities of modern satellites enable efficient data compression, reducing network congestion and latency while supporting near \mbox{real-time information delivery \cite{10185626}.}

These advancements are further reinforced  by the emergence  of \glspl{aeos} \cite{Wang_2021}. While \glspl{ceos} can only adjust their attitude along the roll axis, \glspl{aeos} benefit from full three-axis attitude control (roll, pitch, and yaw), significantly increasing the flexibility of data acquisition by extending the \gls{vtw} -- the time interval during which a specific target can be observed. Consequently, the selection of the actual observation time for a given observation target, referred to as the \gls{otw}, becomes more flexible. This flexibility expands feasible observation schedules but introduces complexity when selecting \glspl{otw} across multiple observation targets within the same \gls{sth}. This is known as the \gls{aeossp}, which aims to maximize the total observation profit for a given set of targets and satellites over the \gls{sth}, while satisfying all temporal, energy, and storage constraints.

The observation profit depends on the context and is commonly defined by target priority and frame quality. The later depends on the relative position between satellite and target during the \gls{otw}, leading to the concept of time-dependent profit. A widely used indicator of frame quality in \gls{eo} is the spatial resolution, typically measured by the \gls{gsd} \cite{ijgi10060406}, which quantifies the actual ground area represented by a single pixel in the frame. As the target deviates from the satellite's nadir point, the \gls{gsd} increases, leading to reduced spatial resolution.

Early research on \gls{aeossp} primarily focused on single-satellite scenarios, which are already challenging. Extending the problem to multi-satellite configurations introduces additional complexity, as each target may be observed by multiple satellites. Solutions are classified into distributed and centralized approaches. In distributed schemes, each satellite operates as an autonomous agent, independently scheduling its own list of targets while coordinating with the rest of satellites in the constellation. Centralized methods, in contrast, rely on a central node --controller-- to assign observation tasks to all satellites in the constellation, sending each satellite a predefined set of instructions to execute \cite{alma991001513474404886}. While distributed strategies are often more scalable and time-efficient, they may yield suboptimal global performance and require constant inter-satellite communication. In contrast, centralized approaches offer globally optimized solutions and are particularly well-suited for small satellite constellations.

In this paper, we present a constructive heuristic algorithm for solving the \gls{aeossp} using a centralized approach tailored to continuous remote monitoring on ground, ensuring high-quality and fresh information of a given set of targets within an edge computing framework that leverages the onboard processing capabilities of satellites. To achieve this, we integrate data processing into the \gls{aeossp} and define the observation profit as a function of both spatial resolution and data freshness, using \gls{gsd} and \gls{aoi} as key performance indicators. The rest of the paper  is organized as follows. We first provide an overview of related works (Section \ref{sec:soa}) and system model (Section \ref{sec:sysmodel}). Then, the timing metrics, optimization problem, priority indicators and algorithm are described in Sections \ref{sec:timing}-\ref{sec:algorithm}. Results and conclusions are presented in Sections \ref{sec:results} and \ref{sec:conclusions}, respectively.

\section{Related work} \label{sec:soa}

In recent years, the \gls{aeossp} has attracted  significant attention from the research community. Existing approaches can be broadly categorized into exact, heuristic, metaheuristic, and machine learning methods. Among these, heuristic and metaheuristic techniques have gained prominence over exact methods due to their ability to provide high-quality solutions with lower computational cost. For example, \cite{XU2016195} propose several priority-based constructive heuristic methods with promising results, while \cite{8750776} addresses the \gls{aeossp} under oversubscribed target scenarios using a feedback structured heuristic, a scenario closely related to our work.

Parallel to advances in scheduling techniques, edge computing has emerged as a key enabler for reducing latency and congestion in satellite network communications. In \cite{10185626}, the authors propose an edge computing framework for real-time, very-high-resolution \gls{eo} imaging, while \cite{VALENTE2023109849} focus on resource allocation strategies in satellite constellations for \gls{eo}. In addition, \cite{10794283} explores satellite edge computing in a real-world application, aiming to design and size a constellation that balances frame quality with information freshness, using \gls{aoi} as a key performance metric. \gls{aoi} is a widely used metric to quantify the timeliness of information, measuring the time elapsed since the last received update \cite{8187436}. If no updates occur, the \gls{aoi} increases linearly; upon receiving a new update, it resets to reflect the time elapsed since the generation of that update. In the context of \gls{eo}, \gls{aoi} represents the time since a particular target was last observed. Over time, various extensions of \gls{aoi} have been proposed, including the \gls{paoi}, which is particularly suited for worst-case optimization scenarios, as it captures the \gls{aoi} value just before an update is received.

\section{System Model} \label{sec:sysmodel}

\begin{figure*}[t]
    \centering
    \begin{subfigure}{0.41\textwidth}
        \centering
        \includegraphics[width=\textwidth]{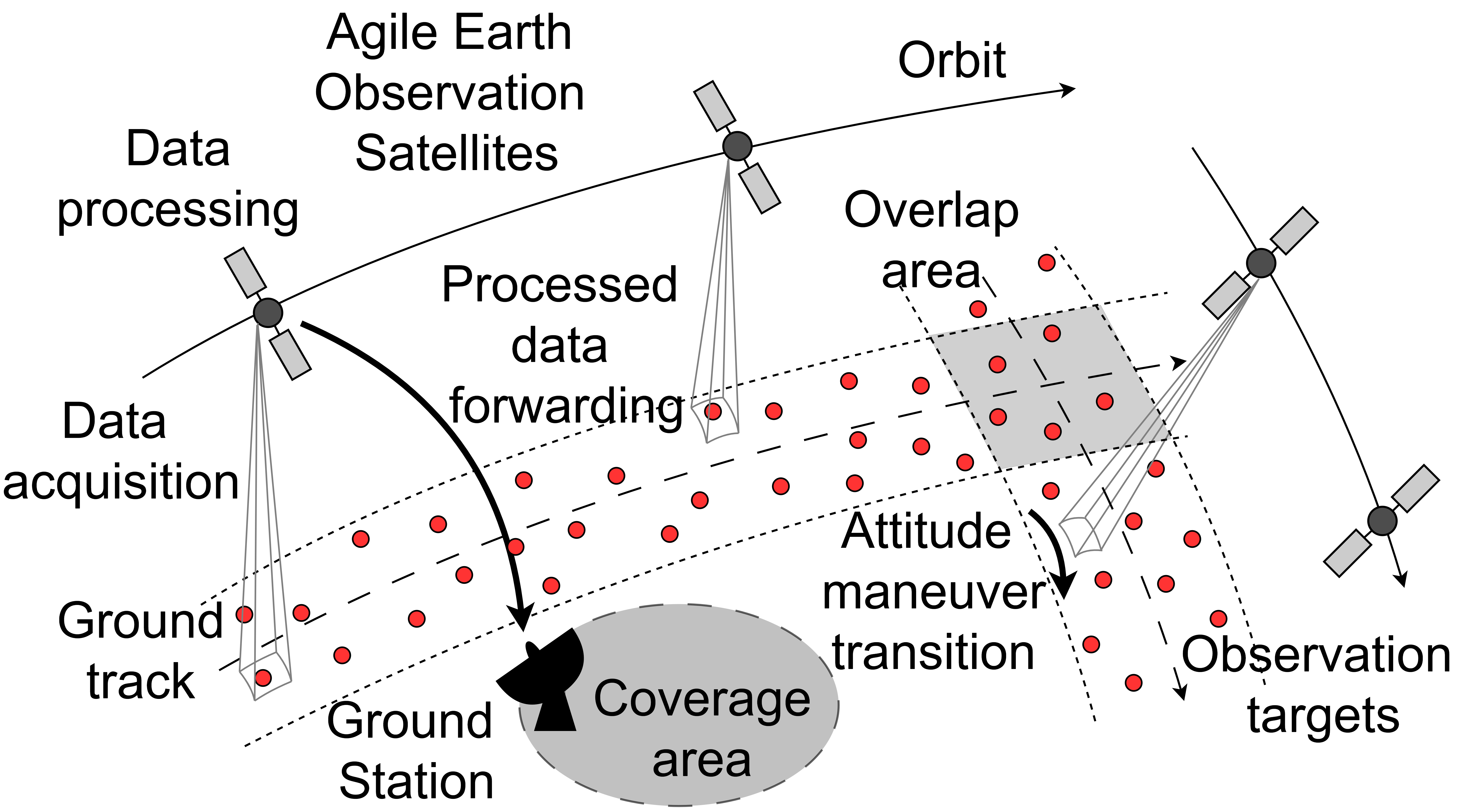}
        \caption{}
        \label{fig:scenario}
    \end{subfigure}
    \hfill
    \begin{subfigure}{0.46\textwidth}
        \centering
        \includegraphics[width=\textwidth]{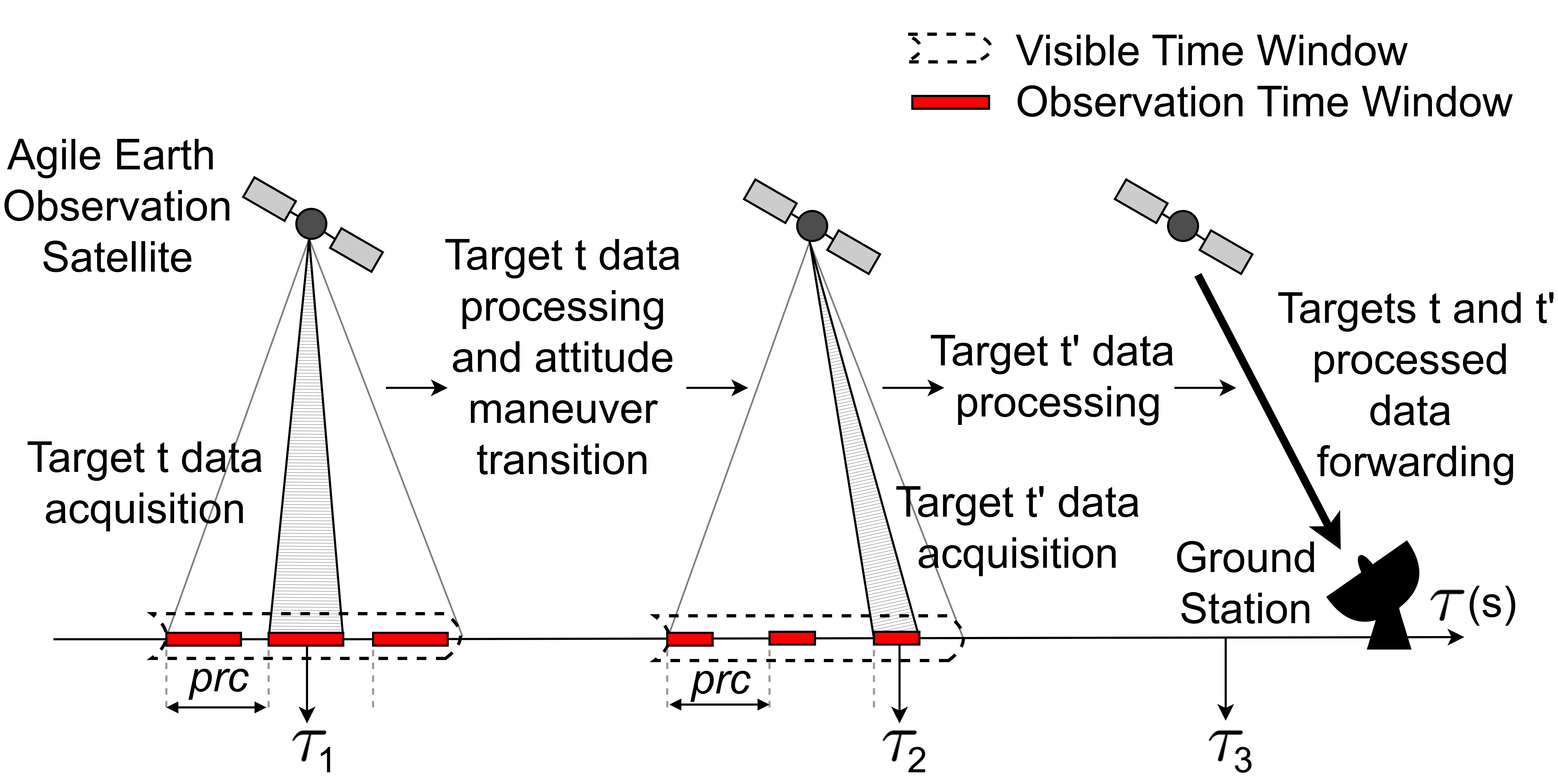}
        \caption{}
        \label{fig:vtw_vs_otw}
    \end{subfigure}
    \hfill
    \caption{(a) Scenario with multiple \glspl{aeos} tasked with observing multiple targets, processing and transmitting the acquired data. (b) Sketch of the task sequence: a satellite observes target $t$ at $\tau_1$ for $\tau_t^{(obs)}$ seconds, processes the acquired data, executes the corresponding attitude maneuvers, observes target $t'$ at $\tau_2$ for $\tau_{t'}^{(obs)}$ seconds, and processes the newly acquired data. The overall processed data is transmitted to ground once a ground station is reachable at $\tau_3$.}
    \label{fig:combined_fig}
    \vspace{-0.4cm}
\end{figure*}

Fig. \ref{fig:scenario} shows the considered scenario, which involves a set of \glspl{aeos} $S$, deployed in a constellation comprising  $N_{OP}$ orbital planes and $N_{sat}$ satellites per plane, operating at an altitude $h$ with an orbital inclination $\iota$. The satellites are arranged according to a Walker-Delta or Walker-Star topology \cite{2022}. These satellites are responsible for observing a set of targets $T$ within a designated observation area, processing the acquired data --i.e. a frame of the target area-- and transmitting the resulting information to the ground segment, represented by a set $G$ of ground stations, when within coverage following a store-and-forward approach (Fig. \ref{fig:vtw_vs_otw}). All the targets in the instance must be observed as regularly as possible. To achieve this, the \gls{sth} is divided into $N_{STP}$ \glspl{stp}, and for each \gls{stp}, every satellite in the deployment receives a new schedule, allowing previous schedules to be taken into account when constructing the current one. Each target $t \in T$ requires an observation duration of $\tau_t^{(obs)}$ seconds and may be observed by a satellite $s \in S$ during multiple orbital passes. Let $O_{s,t}$ denote the set of orbits in which satellite $s$ can observe target $t$. The corresponding \gls{vtw} for each satellite $s$ and target $t$ during orbit $o \in O_{s,t}$ triplet is defined as $VTW_{s,t,o}$ spanning the interval $[sw_{s,t,o}, ew_{s,t,o}]$, where $sw_{s,t,o}$ and $ew_{s,t,o}$ are the start and end times of the \gls{vtw}, respectively. Each $VTW_{s,t,o}$ is further discretized into a set of \glspl{otw} with a fixed time step of $prc$ seconds. Thus, $OTW_{s,t,o,w}$ represents the \gls{otw} for satellite $s$ and target $t$ during orbit $o$ within $VTW_{s,t,o}$, with index $w \in W_{s,t,o}$ where $W_{s,t,o}$ denotes the set of \glspl{otw} contained in $VTW_{s,t,o}$. An $OTW_{s,t,o,w}$ is defined by $[so_{s,t,o,w}, eo_{s,t,o,w}],$ and an associated observation profit $\rho_{s,t,o,w}$, where $so_{s,t,o,w}$ and $eo_{s,t,o,w}$ represent the start and end times of the \gls{otw} according to $\tau^{(obs)}_t$, respectively.

The attitude required for observing  $OTW_{s,t,o,w}$ is defined by its roll ($\theta_{s,t,o,w}$), pitch ($\phi_{s,t,o,w}$), and yaw ($\psi_{s,t,o,w}$) angles. Each satellite's maneuvering capabilities  are restricted by  its maximum roll, pitch, and yaw angles ($\theta_{max}$, $\phi_{max}$, and $\psi_{max}$, respectively), as well as by its attitude transition time. These factors dictate which targets can be observed and scheduled at any given time. The attitude transition time is described by a piecewise linear function that depends on the speed of the camera and the angular difference between two consecutive observations. Specifically, we consider the following attitude transition time between two consecutive observations $OTW_{s,t,o,w}$ and $OTW_{s,t',o,w'}$\cite{LIU201741}:

\vspace{-3mm}
\small
\begin{equation}
    \Delta t_{t,w-t',w'} = 
    \begin{cases}
        11.66, & \alpha_{t,w-t',w'} \leq 10^\circ \\
        5 + \alpha_{t,w-t',w'} / 1.5, & 10^\circ < \alpha_{t,w-t',w'} \leq 30^\circ \\
        10 + \alpha_{t,w-t',w'} / 2, & 30^\circ < \alpha_{t,w-t',w'} \leq 60^\circ \\
        16 + \alpha_{t,w-t',w'} / 2.5, & 60^\circ < \alpha_{t,w-t',w'} \leq 90^\circ \\
        22 + \alpha_{t,w-t',w'} / 3, & \alpha_{t,w-t',w'} > 90^\circ \\
    \end{cases}
    ,
    \label{eq:attitude_transition_time}
\end{equation}
\normalsize
\begin{align}
    \alpha_{t,w-t',w'} = & |\theta_{s,t,o,w} - \theta_{s,t',o,w'}| + |\phi_{s,t,o,w} - \phi_{s,t',o,w'}| \notag \\ & + |\psi_{s,t,o,w} - \psi_{s,t',o,w'}|
    \label{eq:angle_displacement}
    ,
\end{align}
where $\alpha_{t,w-t',w'}$ is the total attitude transition angle between $OTW_{s,t,o,w}$ and $OTW_{s,t',o,w'}$.

Each satellite is equipped with an onboard \gls{cpu} that processes  the collected  data in real time. The processed data, $D'$, is determined by the compression factor $\sigma$, defined as:
\begin{equation}
    D' = \frac{D}{\sigma}
    \label{eq:compression_factor}
    ,
\end{equation}
where $D$ is the amount of data to be processed. The compression factor can range from a few units or tens (traditional compression algorithms) \cite{10185626}; to thousands when using semantic information extraction methods in the context of semantic or goal-oriented communications \cite{10794283}. For the case of a \gls{cpu} (and similarly for a \gls{gpu}), the processing time is calculated as:
\begin{equation}
    \tau^{(proc)}_{s,t,o,w} = \frac{D_{s,t,o,w}C}{N_{cores}f_{CPU}}
    \label{eq:t_proc}
    ,
\end{equation}
where $D_{s,t,o,w}$ is the number of bits to be processed for $OTW_{s,t,o,w}$, $C$ the complexity of the compression algorithm expressed in \gls{cpu} cycles per bit, and $N_{cores}$ and $f_{CPU}$ the number of \gls{cpu} cores and work frequency, respectively. $D_{s,t,o,w}$ is defined as:
\begin{equation}
    D_{s,t,o,w} = \frac{\xi_{s,t,o,w}(\xi_{s,t,o,w} + \tau^{(obs)}_t R_E \omega)}{GSD^2_{s,t,o,w}}q_{px}
    \label{eq:px_to_process}
    ,
\end{equation}
where $\xi_{s,t,o,w}$ is the swath width, i.e, the width of the observable area on the surface of the Earth, determined by the \gls{fov} of the sensor and the relative position between target $t$ and satellite $s$ during $OTW_{s,t,o,w}$; $GSD_{s,t,o,w}$ is the \gls{gsd} for $OTW_{s,t,o,w}$; $R_E$ is the Earth's radius; $\omega$ is the satellite angular velocity; and $q_{px}$ is the bit depth of a pixel in the frame.

Energy consumption for observation, transition, and processing is modeled via $E_{obs}$, $E_{tran}$, and $E_{proc}$, respectively, and all representing energy per time unit. The energy budget for each satellite $s$ during a \gls{stp} is limited by $E_{max}$.

Satellites communicate with ground stations when they are within coverage, taking into account for both propagation and transmission delays. Thus, the communication time is calculated as follows:
\begin{equation}
    \tau_{comm} = {\frac{D_l}{R_l}} + \frac{{d_l}}{c},
    \label{eq:t_comm}
\end{equation}
where $D_l$ is the number of bits to transmit through a given link $l$, $R_l$ the bit rate of such link, $d_l$ the distance between two nodes of the path, and $c$ the speed of light. The time elapsed between data processing and its transmission is denoted as $\tau_{store}$.

\section{Timing metrics} \label{sec:timing}

\begin{figure} [t]
\centering
\includegraphics[width=0.43\textwidth]{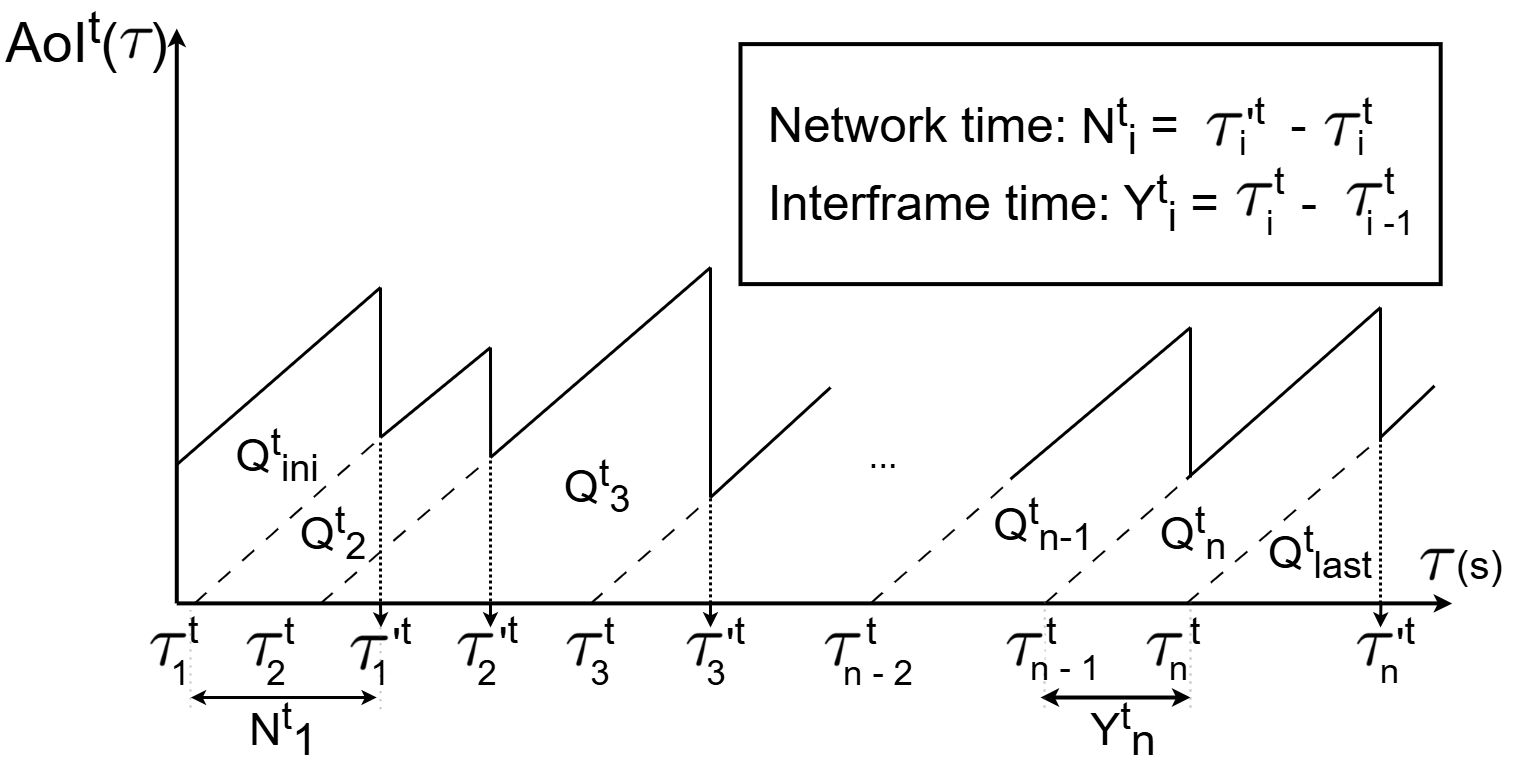}
\caption{Evolution of the \gls{aoi} of target $t$.}
\label{fig:aoi} \vspace{-0.4cm}
\end{figure}

Each observation target $t$ is associated with an \gls{aoi} value denoted as $AoI^{t}$, which accumulates over time until updated, including processing, store, and communication times. In our application, \gls{aoi} is defined as the time elapsed at the ground segment since the last received compressed image of target $t$ was captured by the corresponding satellite\cite{10794283}. Mathematically, this is expressed as follows. Fig. \ref{fig:aoi} shows the time evolution of the \gls{aoi} of a target $t$. It is assumed that the system is first observed at time $\tau = 0$ and index $i$ represents the number of the frame for target $t$. A frame $i$ of target $t$ is captured at $\tau = \tau^{t}_{i}$ and the corresponding compressed information is received at the ground segment at $\tau=\tau'^{t}_{i}$.

$N^{t}_{i}$ is defined as the total network time of the system (processing + store + communication times), $N^t_{i} = \tau'^{t}_{i} - \tau^{t}_{i}$; and $Y^{t}_{i}$ as the interframe time, the time between the capture of two frames for the same target $t$, $Y^{t}_{i} = \tau^{t}_{i} - \tau^{t}_{i - 1}$.

The average \gls{aoi} of target $t$ can be calculated as follows:
\begin{equation}
    \overline{AoI^{t}}=\frac{1}{STH}\left(Q^t_{ini} + Q^t_{last} + \sum_{i = 2}^{N(STH)}{Q^t_{i}}\right);
    \label{eq:aoi_avg}
\end{equation}
where $N(STH)$ is the number of arrivals by the end of the \gls{sth}. Each $Q^t_{i}$ for $1 < i$ is a trapezoid whose area can be calculated as:
\begin{equation}
    Q^t_{i}=\frac{1}{2}\left(N^t_{i} + Y^t_{i}\right)^2 - \frac{1}{2}\left(N_{i}^{t}\right)^{2} = Y^t_{i}N^t_{i} + \frac{\left(Y_{i}^{t}\right)^{2}}{2}.
    \label{eq:q}
\end{equation}
Assuming an initial observation of the system at $\tau = 0$, the average \gls{paoi} can be calculated as:
\begin{equation}
    \overline{PAoI^t}=\frac{1}{N(STH)}\left(\tau'^t_{1} + \sum_{i = 2}^{N(STH)}{(\tau'^{t}_{i} - \tau^t_{i - 1})}\right).
    \label{eq:paoi_avg}
\end{equation}

Furthermore, for our scenario, we define $\delta_t$ as the number of \glspl{stp} that have elapsed since target $t$ was last included in a schedule. This metric is directly related to \gls{aoi}, since targets that have not been observed for a longer time are associated with higher \gls{aoi} values.

\section{Agile Earth Observation Satellite Scheduling Problem} \label{sec:aeossp}

The \gls{aeossp} aims to maximize the total collected observation profit while satisfying all temporal, energy, and storage constraints. In the considered scenario, since data is processed dynamically as it is acquired and transmitted to the ground, storage constraints are neglected, and processing is incorporated into the optimization problem. The mathematical formulation is as follows:

\begin{equation}
    \text{max} \sum_{s \in S} \sum_{t \in T} \sum_{o \in O_{s,t}} \sum_{w \in W_{s,t,o}}  
    \rho_{s,t,o,w} x_{s,t,o,w}
    \label{eq:objective_function}
    ,
\end{equation}
subject to:
\small
\begin{equation}
    sw_{s,t,o} \leq so_{s,t,o,w} + \tau^{(obs)}_t \leq ew_{s,t,o}
    \label{eq:window_constraint}
    ,
\end{equation}
\begin{equation}
    so_{s,t,o,w} + \tau^{(obs)}_t + \text{max}\left(\Delta \tau_{t,w-t',w'}, \tau^{(proc)}_{s,t,o,w}\right) \leq so_{s,t',o,w'}
    \label{eq:attitude_processing_constraint}
    ,
\end{equation}
\begin{align}
     \sum_{t \in T} \sum_{w\in W_{s,t,o}} x_{t,s,o,w} \left( E_{obs}\tau^{(obs)}_t +E_{proc}\tau^{(proc)}_{s,t,o,w}\right) \notag \\ + \sum_{t, t' \in T} \sum_{w, w'\in W_{s,t,o}}y_{t,w,t',w'} \left( E_{tran} \Delta \tau_{t,w-t',w'} \right) \leq E_{max}
     \label{eq:energy_constraint}
     ,
\end{align}
\begin{equation}
     \sum_{s \in S} \sum_{o \in O_{s,t}} \sum_{w \in W_{s,t,o}}{x_{s,t,o,w}} \leq 1
     \label{eq:binary_variable_constraint}
     ,
\end{equation}
\begin{equation}
     x_{s,t,o,w}, y_{t,w,t',w'} \in \{0, 1\}
     \label{eq:binary_variable_value_constraint}
     ,
\end{equation}
\normalsize

\noindent where $x_{s,t,o,w}$ and $y_{t,w,t',w'}$ are binary decision variables, $x_{s,t,o,w} = 1$ denotes that $OTW_{s,t,o,w}$ is scheduled, otherwise $x_{s,t,o,w} = 0$, and $y_{t,w,t',w'} = 1$ that $OTW_{s,t,o,w}$ is scheduled right before $OTW_{s,t',o,w'}$, otherwise $y_{t,w,t',w'} = 0$. Equation (\ref{eq:objective_function}) is the optimization objective function, which is to maximize the sum of collected profits; Equation (\ref{eq:window_constraint}) represents the \gls{vtw} constraint, which ensures that the observation of target $t$ occurs within one of its \gls{vtw}; Equation (\ref{eq:attitude_processing_constraint}) represents the attitude transition and processing time constraint, ensuring that $OTW_{s,t',o,w'}$ can only be scheduled after $OTW_{s,t,o,w}$ if the information associated with $OTW_{s,t,o,w}$ has been fully processed and the attitude maneuver is feasible; Equation (\ref{eq:energy_constraint}) represents the energy constraint, which guarantees that the total energy consumption does not exceed the maximum allowed energy; Equation (\ref{eq:binary_variable_constraint}) denotes that a target can exist at most once in a single schedule; and Equation (\ref{eq:binary_variable_value_constraint}) represents the value of the decision variables.

We consider both the quality of the collected frames, represented by the \gls{gsd}, and the freshness of the last received information, quantified by the \gls{aoi}, when defining the observation profit. The observation profit $\rho_{s,t,o,w}$ ranges from 0 to 1, where 1 indicates that $OTW_{s,t,o,w}$ has the highest possible priority and quality. We formulate it as:
\begin{equation}
    \rho_{s,t,o,w} = \frac{GSD_{nadir}}{GSD_{s,t,o,w}} \frac{\delta_t}{\delta_{max}},
    \label{eq:observation_profit}
\end{equation}
where $GSD_{nadir}$ is the \gls{gsd} at nadir, and $\delta_{max}$ the $\delta_t$ value of the target that has remained unobserved for the longest time in the instance. As mentioned previously, the \gls{gsd} increases as the target moves away from the nadir reaching its minimum at the nadir itself%(see Fig. \ref{fig:gsd_variation})
. A small \gls{gsd} value leads to a better resolution. Thus, the first term of the observation profit refers to spatial resolution. The second term ensures that the targets with higher \gls{aoi} values have higher priority.

\section{Priority Indicators} \label{sec:priority}

When scheduling multiple \gls{aeos}, it is common practice to design priority indicators to ensure that the selected targets in the final schedule yield the highest possible profit.

In our scenario, the primary priority indicator is $\delta_t$, as it directly correlates with the \gls{aoi}. A higher $\delta_t$ value indicates higher priority.

Within the same \gls{stp}, not all targets have the same number of scheduling opportunities, as the number of \glspl{otw} varies for each target in the instance. We define the number of available \gls{otw} for target $t$ as its assignment flexibility \cite{CHEN2018177}, denoted as $FL_t$. A higher $FL_t$ indicates more opportunities for inclusion in the final schedule, whereas a lower $FL_t$ signifies higher priority among targets with the same $\delta_t$ value.

When selecting the \gls{otw} for target $t$, not only its associated observation profit should be considered, but also the observation profit of conflicting \glspl{otw} that would be excluded from the schedule. Two \glspl{otw} are considered to be in conflict when, disregarding the energy constraint (Equation \ref{eq:energy_constraint}), scheduling both would render the optimization problem infeasible. This concept is well captured by the \gls{oc} \cite{XU2016195},  which is mathematically defined as the sum of the observation profits of the conflicting \glspl{otw}:
\begin{equation}
    OC_{s,t,o,w} = \sum_{t',w' \in COTW_{s,t,o,w}}{\rho_{s,t',o,w'}},
    \label{eq:operation_cost}
\end{equation}
where $COTW_{s,t,o,w}$ represents the set of \glspl{otw} that conflict with $OTW_{s,t,o,w}$. A lower $OC_{s,t,o,w}$ value signifies higher priority among the \glspl{otw} for the same target.

\section{Algorithm} \label{sec:algorithm}

Based on the priority indicators introduced in Section \ref{sec:priority}, we propose a constructive heuristic algorithm to solve the \gls{aeossp} for continuous monitoring. For each \gls{stp}, the targets are grouped based on $\delta_t$, with groups arranged in descending order of $\delta_t$. Within each group, targets are ranked in ascending order of $FL_t$, and their \glspl{otw} are sorted in ascending order of \gls{oc}. The procedure is carried out as follows: for each target $t$, the first available $OTW_{s,t,o,w}$ is checked against the scheduling constraints. If feasible, it is added to the schedule, and the next target $t'$ is considered; otherwise, the next \gls{otw} for the same target is evaluated until either one is scheduled or all available \glspl{otw} are exhausted. This process is repeated for every group until all targets have been considered.

While constructive heuristics provide fast solutions, they can also be suboptimal, as they rely on greedy decision-making without exploring alternative configurations that could yield better overall results. For this reason, \gls{ls} \cite{XU2016195} or feedback \cite{8750776} strategies are applied to refine the initial solution. We propose a \gls{ls} strategy in which new solutions are generated following an insertion and removal policy, i.e., attempting to schedule previously unscheduled targets by removing some of the scheduled ones. The procedure unfolds as follows:

\noindent \textbf{Step 1:} Unscheduled targets and their corresponding \glspl{otw} are listed. The set of unscheduled targets is denoted as $U$. The \gls{oc} for every listed \gls{otw} is computed accounting only for the conflicting missions in the current schedule. Both targets and \glspl{otw} are then arranged following the same priority policies used in the constructive heuristic.

\noindent \textbf{Step 2:} Let $u$ and $OTW_{s,u,o,w}$ be the current unscheduled target and \gls{otw}, respectively. If temporal constraints are met, proceed to Step 3; otherwise, check wether $OC_{s,u,o,w}$ is lower than $\rho_{s,u,o,w}$. If so, provisionally remove the conflicting scheduled missions and move to Step 3; otherwise consider the next \gls{otw} for $u$ or, if no more \glspl{otw} are available for target $u$, the next unscheduled target $u'$ and repeat Step 2.

\noindent \textbf{Step 3:} Check energy constraints. If the schedule is feasible, add $OTW_{s,u,o,w}$ to the schedule and consider the next unscheduled target $u'$; otherwise, remove the scheduled \glspl{otw} with lower observation profit from the schedule of satellite $s$ until the optimization problem is feasible and add provisionally $OTW_{s,u,o,w}$. If the new solution improves upon the current one, update the schedule; otherwise, consider the next \gls{otw} for $u$ or, if no more \glspl{otw} are available for target $u$, the next unscheduled target $u'$ and go to Step 2.

This process is repeated until all unscheduled targets have been considered. The pseudocode of the proposed algorithm is deployed in Algorithm \ref{alg:algorithm}.

\begin{figure}[t]
\begin{algorithm}[H]
\caption{Constructive Heuristic Algorithm}
\begin{algorithmic}[1]
\STATE {\bfseries Input:} $S$, $T$, $G$, $STH$, $N_{STP}$
\STATE {\bfseries Output:} Schedule
\STATE Initialize Schedule $\gets \varnothing$
\STATE Calculate \glspl{otw}.
\FOR{$STP$ in $STH$}
\STATE Calculate priority indicators $\delta$, FL, and \gls{oc}.
\STATE Sort targets according to the proposed priority policies.
\FOR{$t$ in $T$}
\STATE Sort \glspl{otw} for $t$ and current $STP$, $OTW_{t, STP}$, in ascending \gls{oc} order.
\WHILE{$t$ is not scheduled \AND $OTW_{t, STP}$ is not empty}
\STATE $OTW_{s,t,o,w} \gets OTW_{t, STP}[0]$
\IF{scheduling $OTW_{s,t,o,w}$ is feasible}
    \STATE Add $OTW_{s,t,o,w}$ to Schedule.
\ELSE
    \STATE Remove $OTW_{s,t,o,w}$ from $OTW_{t, STP}$.
\ENDIF
\ENDWHILE
\ENDFOR
\STATE List unscheduled targets $U$.
\FOR{$u$ in $U$}
\STATE Sort \glspl{otw} for $u$ and current $STP$, $OTW_{u, STP}$, in ascending \gls{oc} order accounting only for the conflicting missions in Schedule.
\WHILE{$u$ is not scheduled \AND $OTW_{u, STP}$ is not empty}
\STATE $OTW_{s,u,o,w} \gets OTW_{u, STP}[0]$.
\STATE Construct feasible solution, Schedule', scheduling $OTW_{s,u,o,w}$ following the proposed insertion and removal policy.
\IF{Schedule' improves Schedule}
    \STATE $\text{Schedule} \gets \text{Schedule'}$
\ELSE
    \STATE Remove $OTW_{s,u,o,w}$ from $OTW_{u, STP}$.
\ENDIF
\ENDWHILE
\ENDFOR
\STATE Update \gls{aoi} for every target $t \in T$
\ENDFOR
\end{algorithmic}
\label{alg:algorithm}
\end{algorithm}
\vspace{-2.5em}
\end{figure}

\section{Results} \label{sec:results}

The simulation parameters are as follows. The space segment is composed of $N_{op}=4$ orbital planes and $N_{sat}=2$ satellites per orbital plane, deployed at $h = 600$ kilometers, with an orbital inclination $\iota = 53^\circ$, distributed according to a walker delta topology. Each satellite owns a \gls{cpu} with $N_{cores}=8$ and $f_{CPU} = 1.8$ GHz, and a camera whose attitude maneuvers are limited by  $\theta_{\max} = 45^\circ$, \mbox{$\phi_{\max} = 45^\circ$}, and $\psi_{\max} = 90^\circ$ that allows a $GSD_{nadir} = 0.5$ m/pixel and a swath from $\xi = 5$ to $\xi = 7.29$ kilometers. The bit depth of the captured frames is set to $q_{px} = 11$ bits/pixel. $E_{obs}$, $E_{proc}$, and $E_{tran}$ values are set to 2 units/second, while $E_{max}=5000$ units. We consider a complexity of the compression algorithm of $C = 100$ CPU cycles per bit and a compression factor of $\sigma = 10$ is assumed. The KSAT Global Ground Station network is regarded as the ground segment. We consider a \gls{sth} of 10 orbital periods and $N_{STP} =10$, so that each orbital period corresponds to one \gls{stp}, and instances of $\{1000, 1200, 1400, 1600, 1800\}$ targets uniformly distributed across the coverage area of the satellites. Each target must be observed for a duration $\tau^{(obs)}_t = Uniform(1,5)$ seconds, and a discretization step of $prc = 10$ seconds is assumed.

The performance of the proposed algorithm will be compared with a \gls{fifo} constructive heuristicc, which schedules feasible targets in ascending \gls{vtw} order, capturing them at the first feasible \gls{otw}.

\begin{figure} [t]
\centering
\includegraphics[width=0.445\textwidth]{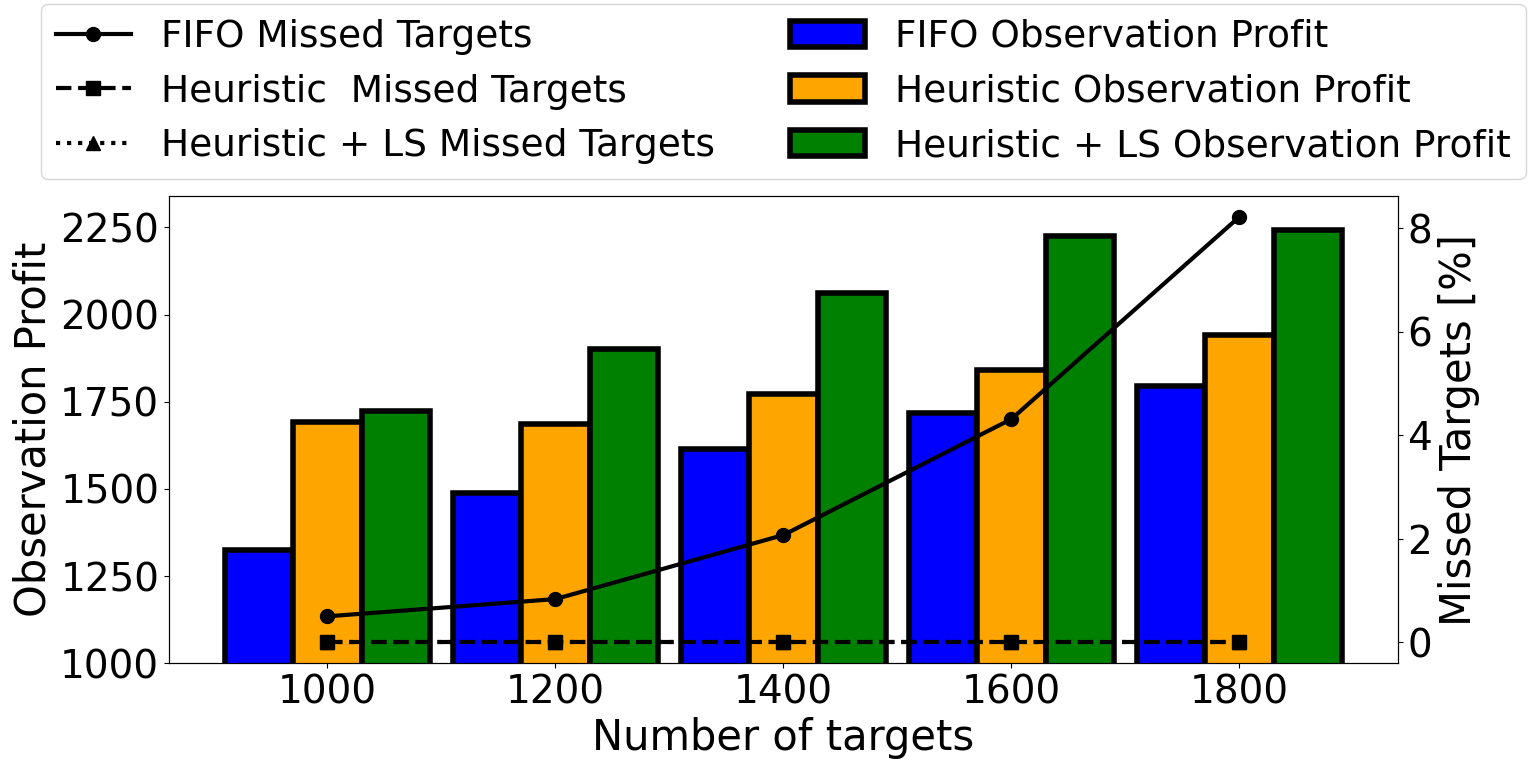}
\caption{Total collected observation profit by the end of the \gls{sth} for all the instances.}
\label{fig:observation_profit} \vspace{-0.4cm}
\end{figure}

\begin{figure} [t]
\centering
\includegraphics[width=0.445\textwidth]{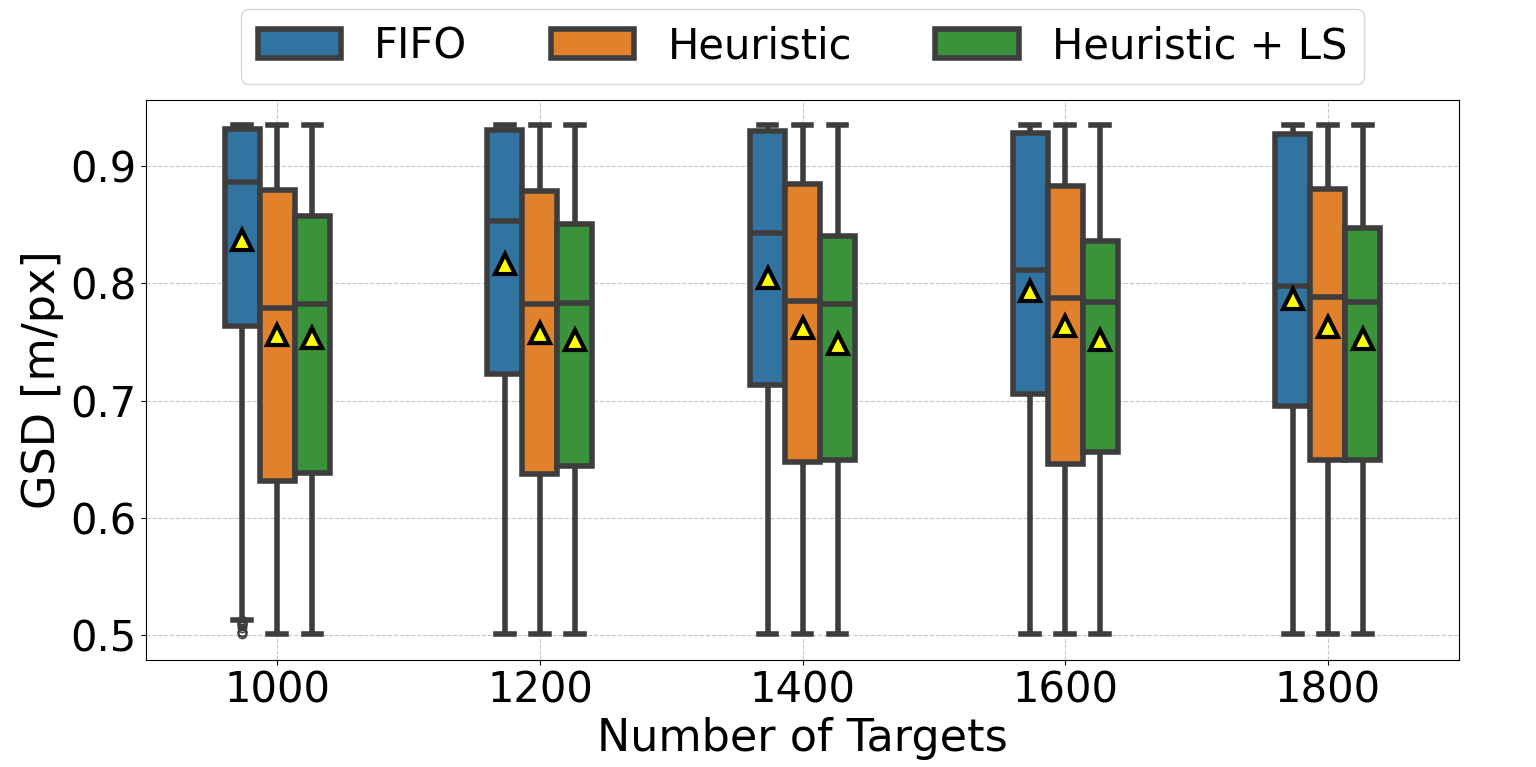}
\caption{Boxplot of the \gls{gsd} of the captured frames by the end of the \gls{sth}.}
\label{fig:results_gsd} \vspace{-0.4cm}
\end{figure}

\begin{figure} [t]
\centering
\includegraphics[width=0.445\textwidth]{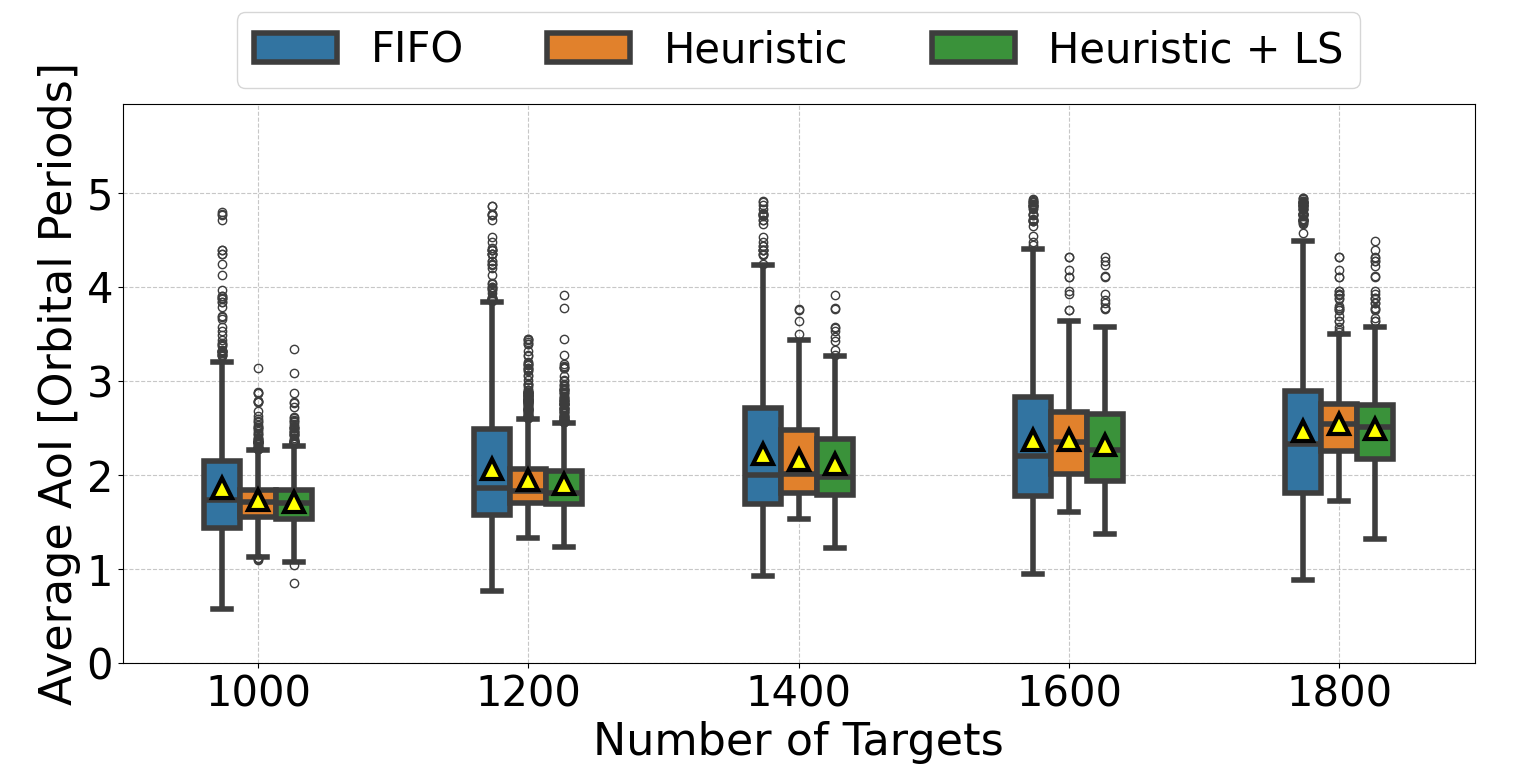}
\caption{Boxplot of the average \gls{aoi} of in the targets in the instance that have been captured by the end of the \gls{sth}.}
\label{fig:results_aoi_avg} \vspace{-0.4cm}
\end{figure}

\begin{figure} [t]
\centering
\includegraphics[width=0.445\textwidth]{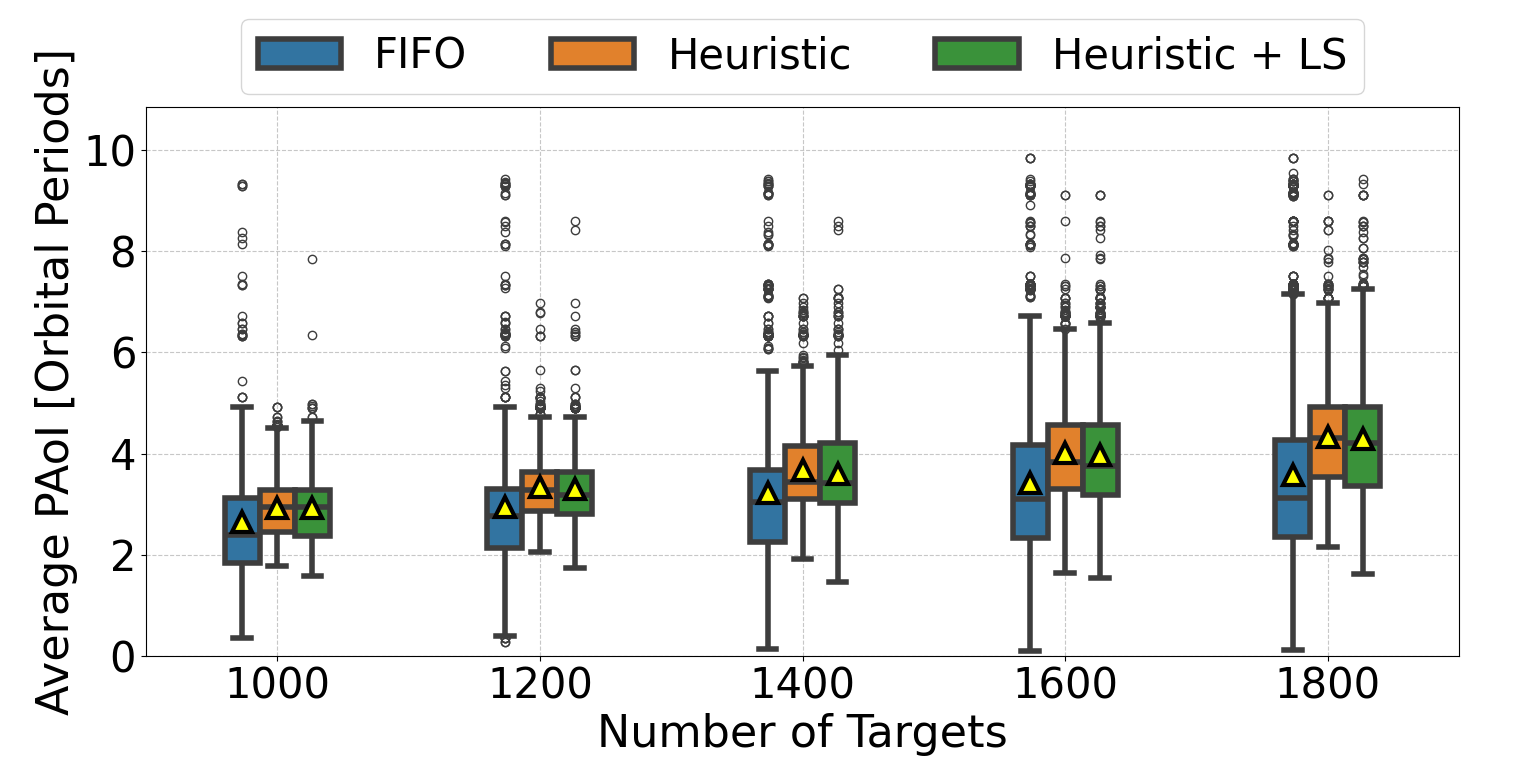}
\caption{Boxplot of the average \gls{paoi} of the targets in the instance that have been captured by the end of the \gls{sth}.}
\label{fig:results_paoi_avg} \vspace{-0.4cm}
\end{figure}

Fig. \ref{fig:observation_profit} shows the total observation profit and the percentage of missed targets by the end of the \gls{sth}. The proposed constructive heuristic algorithm outperforms \gls{fifo} algorithm in all the considered instances. While \gls{fifo} fails to observe all targets, increasing the proportion of missed targets as the number of targets grows, the proposed method (with and without \gls{ls}) monitors all targets, making their curves overlap in the figure. Moreover, the application of the \gls{ls} strategy further enhances the initial solution. These impressions are reinforced when analyzing the results in more detail. Fig. \ref{fig:results_gsd} presents the boxplot of the \gls{gsd} of the captured frames, highlighting that the proposed method achieves higher-quality frames compared to \gls{fifo} approach, with a  \gls{gsd} reduction of up to 10\% on average. Fig. \ref{fig:results_aoi_avg} displays the boxplot of the average \gls{aoi} of the monitored targets by the end of the \gls{sth}. The solution of the proposed method exhibits lower variance, indicating that most targets are observed with similar regularity. Specifically, a variance reduction of up to 83 \% is achieved. This trend is also noticed in Fig. \ref{fig:results_paoi_avg}, which presents the boxplot of the average \gls{paoi} for the monitored targets. Since \gls{fifo} method does not account for information freshness, it often schedules recently observed targets, whereas the heuristic approach prioritizes targets that have remained unobserved for longer periods. This is reflected in the 99th percentile values, which range from 7.33 to 9.34 orbital periods for \gls{fifo}, while the heuristic approach achieves lower values, ranging from 4.56 to 7.32 without \gls{ls} and from 4.65 to 7.86 with \gls{ls}.

Thus, we can conclude that the proposed method for solving the \gls{aeossp} can provide high-quality and up-to-date information of all the targets in the instance in a multi-satellite, continuous, and real-time monitoring scenario.

\section{Conclusions} \label{sec:conclusions}

In this paper, we propose a constructive heuristic algorithm, further enhanced with a \gls{ls} strategy, to solve the \gls{aeossp} in a multi-satellite, continuous, and real-time monitoring scenario. Onboard data processing has been integrated into the optimization problem and the observation profit has been defined in order to consider both image quality and information freshness, using \gls{gsd} and \gls{aoi} as key performance indicators. The obtained results demonstrate that the proposed approach can collect quality frames while ensuring regular monitoring of most targets in the instance. Future work will address weather uncertainty and refine the use of the proposed priority indicators to develop more advanced solutions.

\bibliographystyle{IEEEtran}

\bibliography{refs}

\end{document}